\documentclass[conference]{IEEEtran}
\IEEEoverridecommandlockouts

\usepackage{cite}
\usepackage{amsmath,amssymb,amsfonts}
\usepackage{algorithmic}
\usepackage{graphicx}
\usepackage{textcomp}
\usepackage{xcolor}
\usepackage{caption}
\usepackage{url}
\usepackage{ifthen}
\usepackage{cite}
\usepackage{bbold}
\usepackage{bookmark}
\usepackage{tikz}
\usetikzlibrary{shapes.geometric, arrows, positioning}

\usepackage[capitalise]{cleveref}
\usepackage{balance}
\usepackage{amsthm}
\usepackage{amssymb}
\usepackage{multirow}
\usepackage{amsfonts}
\usepackage{graphicx}
\usepackage{algorithm}
\usepackage{url}
\usepackage{lipsum}
\usepackage{booktabs,makecell, multirow}
\usepackage{tikz}
\usetikzlibrary{matrix}
\usepackage{rotating}
\usepackage{mathtools}
\usepackage{tkz-tab}

\usepackage{latexsym}
\usepackage{color}
\usepackage{graphicx}
\usepackage{threeparttable}
\usepackage{float}
\usepackage{orcidlink}

\usepackage{url}

\usepackage{chngpage}

\usepackage{tabularx}%

\usepackage{booktabs}
\usepackage{pbox}

\usepackage{caption,subcaption}

\usepackage{mathtools}
\usepackage{xcolor}         

\usepackage{titletoc}
\usepackage{titlesec}
\usepackage{hyperref}

\errorcontextlines\maxdimen

\newcounter{algsubstate}

\DeclareGraphicsExtensions{.tif,.pdf,.jpeg,.png,.jpg,.bmp,.svg,.eps,.EMF}

\newtheorem{theorem}{Theorem}
\newtheorem{definition}{Definition}
\newtheorem{lemma}{Lemma}

\hypersetup{
  colorlinks=false, 
  pdfborder={0 0 1} 
}

\newcommand{\localcref}[1]{%
  \hyperref[#1]{%
    \llap{\hbox to 10mm{\hfill}}%
    \cref{#1}%
    \rlap{\hbox to -2mm{\hfill}}%
  }%
}

\def\BibTeX{{\rm B\kern-.05em{\sc i\kern-.025em b}\kern-.08em
    T\kern-.1667em\lower.7ex\hbox{E}\kern-.125emX}}
\begin{document}

\title{Improved Constants in the Kautz–Singleton Construction for Probabilistic Group Testing}

\author{\IEEEauthorblockN{Trung-Khang Tran}
\IEEEauthorblockA{{National University of Singapore}}
\and
\IEEEauthorblockN{Jonathan Scarlett}
\IEEEauthorblockA{{National University of Singapore}}
}

\maketitle

\begin{abstract}
We consider the probabilistic nonadaptive group testing problem, in which a uniformly random defective set of size $d$ is to be identified among $N$ items using a binary test matrix. For the explicit Kautz--Singleton construction, it is known that order-optimal scaling $T=\Theta(d\log N)$ for high-probability recovery can be achieved in the regime $d=\Omega(\log^2 N)$. However, the leading constant in this scaling remains uncharacterized even for standard low-complexity decoders. In this work, we revisit the probabilistic analysis of this construction and improve the leading constant for the COMP decoder, as well as establishing the same bound for the DD decoder. Our approach relies on a sharper analysis of the number of distinct roots associated with the defective items, yielding tighter bounds on the decoding error probability. In particular, we show that both COMP and DD achieve vanishing average error probability with $T\le (1/\ln 2+\varepsilon)d\log N$ tests for any $\varepsilon>0$, in the admissible regime $N=q^k$ for some prime power $q = \Theta(d)$ and integer $k \ge 2$.  For COMP, this leading constant matches that of the random near-constant column weight design, whereas for DD, a gap remains except in the sparse regime $d = N^{o(1)}$.

\end{abstract}

\section{Introduction}
Group testing, originally introduced by Robert Dorfman in 1943 \cite{dorfman1943detection} to screen World War II draftees for syphilis, provides an efficient alternative to individual testing. Given a population of $N$ items containing a small set of $d$ defective items, the objective is to identify the defectives by performing binary tests on groups of items. Each test returns a positive outcome if the tested group contains at least one defective item, and a negative outcome otherwise. By testing items collectively, group testing can identify the defective set using substantially fewer than $N$ tests when $d \ll N$. Since its introduction, group testing has found applications across various fields, including biology \cite{chen2008survey,farach1997group,knill1995group}, medicine \cite{ganesan2017learning,tan2020considerations}, machine learning \cite{malioutov2013exact,augenblick2020group}, data analysis \cite{delaigle2011nonparametric,manoli2006group}, signal processing \cite{gilbert2008group,gilbert2012recovering,cheraghchi2020combinatorial,li2026robust}, and wireless communications \cite{berger1984random,wolf1985born,luo2008neighbor,fletcher2009sparsity}.  We focus on \emph{non-adaptive testing}, in which all tests must be chosen prior to observing any outcomes, which is often highly desired due to reduced overhead and allowing testing in parallel.

Group testing schemes can be classified into two main settings based on their recovery guarantees, namely \textit{combinatorial} \cite{ngo2000survey,du1999combinatorial} and \textit{probabilistic} \cite{atia2012boolean,sejdinovic2010note,chan2011non,scarlett2016phase,aldridge2014group,johnson2018performance,scarlett2018near,scarlett2020noisy} group testing. In the probabilistic setting, the defective set of size $d$ is drawn randomly, and an arbitrarily small probability of reconstruction error is allowed. In this setting, the optimal number of tests scales as $T=\Theta(\min(N,d\log N))$ \cite{bay2022optimal}, and particular attention has been paid to the regime $d=\Theta(N^{1-\theta})$ for some $\theta\in(0,1)$ \cite{chan2011non,scarlett2016phase,aldridge2014group}. In contrast, combinatorial group testing requires zero-error recovery of any defective set containing at most $d$ items, for which at least $T=\min\{N,\Omega(d^2\log_d N)\}$ tests are necessary \cite{d1982bounds,furedi1996onr}. A classical explicit construction of the test matrix was introduced by Kautz and Singleton \cite{kautz1964nonrandom}, achieving $T=O(d^2\log_d^2 N)$, while a subsequent construction by Porat and Rothschild \cite{porat2008explicit} achieves $T=O(d^2\log N)$. Notably, the Kautz–Singleton construction matches the best known lower bound when $d=\Theta(N^\theta)$ for some $\theta\in(0,1)$, but is suboptimal for moderate values of $d$, such as $d=O(\mathrm{poly}(\log N))$, where the Porat–Rothschild construction is more efficient. 


In probabilistic group testing, it is well established that the order-optimal scaling $T=O(d\log N)$ can be achieved using randomized test designs \cite{aldridge2014group,aldridge2017capacity,aldridge2016improved,johnson2018performance}. Beyond this asymptotic order, considerable attention has been devoted to characterizing the leading constant, particularly under low-complexity decoding algorithms. For the COMP decoder, random Bernoulli test designs achieve a leading constant of $e\ln 2 \approx 1.884$, while random near-constant-column-weight designs improve this constant to $1/\ln 2 \approx 1.443$ \cite{aldridge2016improved,johnson2018performance}. The latter constant also matches that of DD when $d = N^{o(1)}$, but DD has a better constant when $d = \Theta(N^{\theta})$.  

Other prior works improved these constants further using more sophisticated decoders \cite{scarlett2016phase,coja2020information,coja2020optimal}, culminating in an information-theoretically optimal threshold that can be attained with polynomial-time decoding (using a random spatially-coupled design) \cite{coja2020optimal}.  Since our focus is on the simpler COMP and DD decoders, we skip the details of these advancements and refer the interested reader to \cite{aldridge2026group}

For explicit test designs, significantly less is known compared to randomized designs.  The work of Inan \emph{et al.}~\cite{inan2019optimality} established that the Kautz--Singleton construction achieves the order-optimal scaling $T=O(d\log N)$ with vanishing average error probability when $d=\Omega(\log^2 N)$. However, their analysis yields a leading constant of $4$, i.e., $T\le (4+o(1))d\log N$, as discussed further in Section~\ref{subsec: Asym for KS}. Thus, despite achieving the optimal asymptotic order, a substantial gap remains between the known leading constants for explicit and randomized test designs.

Motivated by this gap, we revisit the probabilistic analysis of the Kautz--Singleton construction and sharpen the leading constant by deriving a tighter concentration bound on the number of distinct roots of randomly selected Reed--Solomon polynomials. In particular, we show that, for any $\varepsilon>0$, the Kautz--Singleton construction achieves vanishing average error probability with $T\le (1/\ln 2+\varepsilon)d\log N$ tests under both COMP and DD decoding, for admissible parameters (Definition~\ref{def:admissible}). For COMP, this matches the best known constant achieved by random near-constant-column-weight designs, while for DD, it matches the corresponding random-design constant when $d = N^{o(1)}$. 
\section{Problem Setup and KS Construction}\label{sec: Prelim}

Throughout, $\log$ denotes the base-$2$ logarithm and $\ln$ denotes the natural logarithm.  Asymptotic statements are with respect to $N \to \infty$, where other problem parameters (e.g., $d$) may depend on $N$.

Let $A\in\{0,1\}^{T\times N}$ denote the test matrix, whose $N$ columns correspond to the items and $T$ rows correspond to the tests. Specifically, we set $A_{ij}=1$ if item $j$ participates in test $i$, and denote the $j$-th column of $A$ by $A_j$. For convenience, we write $[m]:=\{1,\ldots,m\}$ for any positive integer $m$.

The defective set $S\subseteq[N]$, with $|S|=d$, is drawn uniformly at random from all $d$-subsets of $[N]$. A test outcome is positive if and only if the test contains at least one defective item. Accordingly, the vector of test outcomes is given by
\begin{align}
Y
=
\bigvee_{j\in S} A_j,
\end{align}
where $\vee$ denotes the coordinate-wise Boolean OR. Given the test matrix $A$ and the observation vector $Y$, a decoder produces an estimate $\widehat S$ of the unknown defective set $S$.

We focus on two standard low-complexity decoding algorithms, namely Combinatorial Orthogonal Matching Pursuit (COMP) and Definite Defectives (DD):
\begin{itemize}
    \item The \emph{COMP decoder} first eliminates every item that appears in at least one negative test and then declares all remaining items defective.
    \item  The \emph{DD decoder} begins with the same elimination step, thereby forming a set of possible defectives, and then declares an item defective if it is the unique possible defective appearing in some positive test. 
\end{itemize}
We consider exact recovery, and measure performance through the average probability of error:
\begin{align}
P_e
:=
\Pr(\widehat S\neq S).
\end{align}
The objective is to design explicit test matrices for which the number of tests $T$ is as small as possible while still ensuring that $P_e\to0$ as $N\to\infty$.  Formally, the construction that we consider falls in the following class \cite{inan2019optimality}.

\begin{definition}
A test matrix is \emph{strongly explicit} if any given entry can be computed deterministically in ${\rm poly}(\log N, \log T)$ time given the row and column indices.
\end{definition}

Note that the randomness in our error probability comes from the defective set, and not the matrix construction.

\subsection{Kautz-Singleton Construction} \label{sec:KS}

Here we describe the Kautz-Singleton construction, which was introduced for combinatorial group testing in \cite{kautz1964nonrandom}, and adapted to probabilistic group testing in \cite{inan2019optimality}.

Let $\mathbb F_q$ be a finite field, and let $\alpha_1,\ldots,\alpha_n\in\mathbb F_q$ be distinct evaluation points with $k\le n\le q$. A Reed--Solomon code of dimension $k$ over $\mathbb F_q$ associates each message $m=(m_0,\ldots,m_{k-1})\in\mathbb F_q^k$ with the polynomial
\begin{align}
f_m(x)
=
\sum_{\ell=0}^{k-1}m_\ell x^\ell
\end{align}
and encodes it through evaluations at $\alpha_1,\ldots,\alpha_n$:
\begin{align}
\operatorname{RS}(m)
=
\bigl(
f_m(\alpha_1),\ldots,f_m(\alpha_n)
\bigr).
\end{align}
The Kautz--Singleton construction starts with $\operatorname{RS}(m)$ and replaces each symbol in $\mathbb F_q$ by its corresponding unit vector in $\{0,1\}^q$. Applying this mapping coordinate-wise to all $N=q^k$ Reed--Solomon codewords produces a binary test matrix with $T=nq$ rows. Equivalently, the tests may be indexed by pairs $(a,y)\in\Psi\times\mathbb F_q$, where $\Psi=\{\alpha_1,\ldots,\alpha_n\}$, and the item associated with a polynomial $f$ participates in test $(a,y)$ if and only if $f(a)=y$. Given the field representation and evaluation points, computing $f(a)$ and comparing it with $y$ computes an entry of $A$ in time polynomial in $k\log q=\log N$, so this construction is strongly explicit. 

For brevity, we will often refer to \emph{admissible} parameters to mean the following.

\begin{definition}[Admissible parameters]\label{def:admissible}
A sequence $(q,d,k,N)$ is \emph{admissible} if $q\to\infty$ ranges over prime powers, $d\ge1$ and $k\ge2$ are integers, $N=q^k$, and $c_N:=\frac{q}{d} \to c$ 
for some fixed constant $c > 0$.
\end{definition}

We will sometimes abuse notation and write $q=cd$ with the implicit understanding that $q/d=c+o(1)$ along such admissible prime-power sequences.
As in \cite{inan2019optimality}, our main results will also assume $d=\Omega(\log^2N)$ and choose $c$ as part of the construction.
We observe that for fixed $k \ge 2$, admissible parameters have $d=\Theta(N^{\theta})$ with $\theta = 1/k$, while growing $k$ corresponds to $d=N^{o(1)}$.  In particular, our results (and those in \cite{inan2019optimality}) do not cover an arbitrary prescribed exponent $\theta\in(0,1)$; these could conceivably be handled by shortening the Reed–Solomon code to retain only the required number of items, but it appears to be non-trivial to extend the probabilistic analysis to this scenario.

\subsection{Existing Bound for the Kautz--Singleton Construction}\label{subsec: Asym for KS}
Inan \emph{et al.}~\cite{inan2019optimality} established an order-optimal probabilistic recovery guarantee for the Kautz--Singleton construction under the cover decoder in the regime $d=\Omega(\log^2 N)$. In particular, they showed that an appropriate choice of the Reed--Solomon parameters yields a vanishing average probability of error using $T=\Theta(d\log N)$ tests. For completeness, we restate their result below in the notation used throughout this paper.

\begin{theorem}\label{thm: asymBound}
	\emph{\cite{inan2019optimality}}
	Consider an admissible sequence (Definition~\ref{def:admissible}) with $q/d\to c \ge 4$ and $d=\Omega(\log^2 N)$. For any fixed $\delta>0$, the Kautz--Singleton construction with $n=\lceil(1+\delta)\log N\rceil$ achieves an average probability of error $P_e\le N^{-\Omega(\log q)}+N^{-\delta}$ under the COMP decoder.
\end{theorem}

This result, along with our own results to follow, is illustrated in Figure \ref{fig:KS-rates}, where we adopt the widely-used convention ${\rm rate} = \frac{1}{T} \log_2 \binom{N}{d}$ (bits/test) \cite{aldridge2026group}.


\section{Improved Constants for the KS Construction}\label{sec:improved}

\subsection{Warm-up: A Straightforward Refinement}
From Theorem~\ref{thm: asymBound}, if we take $c=4$ and let $\delta>0$ be arbitrarily close to $0$, we obtain the upper bound $$T\le (4+o(1))d\log N$$ on the number of tests.  Our first result is based on a straightforward refinement of their analysis that improves this constant.

\begin{theorem}
\label{thm: warmup}
For every $\varepsilon>0$, there is a fixed $c>0$ such that, along any admissible sequence (Definition~\ref{def:admissible}) with $q/d\to c$ and $d=\Omega(\log^2N)$, the Kautz--Singleton block length $n$ can be chosen such that COMP recovers a uniformly random defective set of size $d$ with the number of tests satisfying
\begin{equation}
T \le (e\ln 2+\varepsilon)d\log N,
\end{equation}
while the average probability of error satisfies $P_e \to 0$ as $N\to\infty$. (Numerically, $e\ln2\approx1.8842$.)
\end{theorem}
\begin{proof}[Proof sketch]
We follow the proof of Theorem~1 in~\cite{inan2019optimality} and retain its main probabilistic argument. In particular, the COMP error event is reduced to the event that the root set of a product of $d$ randomly selected Reed--Solomon polynomials contains all $n$ evaluation points. The improvement comes from avoiding the specific parameter choices made in~\cite{inan2019optimality}. There, the number of distinct roots is controlled using the fixed threshold $2d$, after which the alphabet size is set to $q=4d$. We instead introduce a general threshold $(1+\eta)d$ and assume $q/d\to c$. The same concentration argument as \cite{inan2019optimality} shows that exceeding the new threshold still has vanishing probability. Conditioned on staying below this threshold, the conditional-uniformity argument of~\cite{inan2019optimality} gives a corresponding upper bound on the probability that all evaluation points are covered.

Keeping $c$ and $\eta$ free therefore yields a family of valid bounds rather than fixing them a priori. Optimizing these parameters, together with the Reed--Solomon block length $n$, gives the leading constant $e\ln 2$. The full proof is provided in Appendix~\ref{app: proofthmwarmup}. 
\end{proof}

\begin{figure}[!t]
	\centering
	\includegraphics[width=\columnwidth]{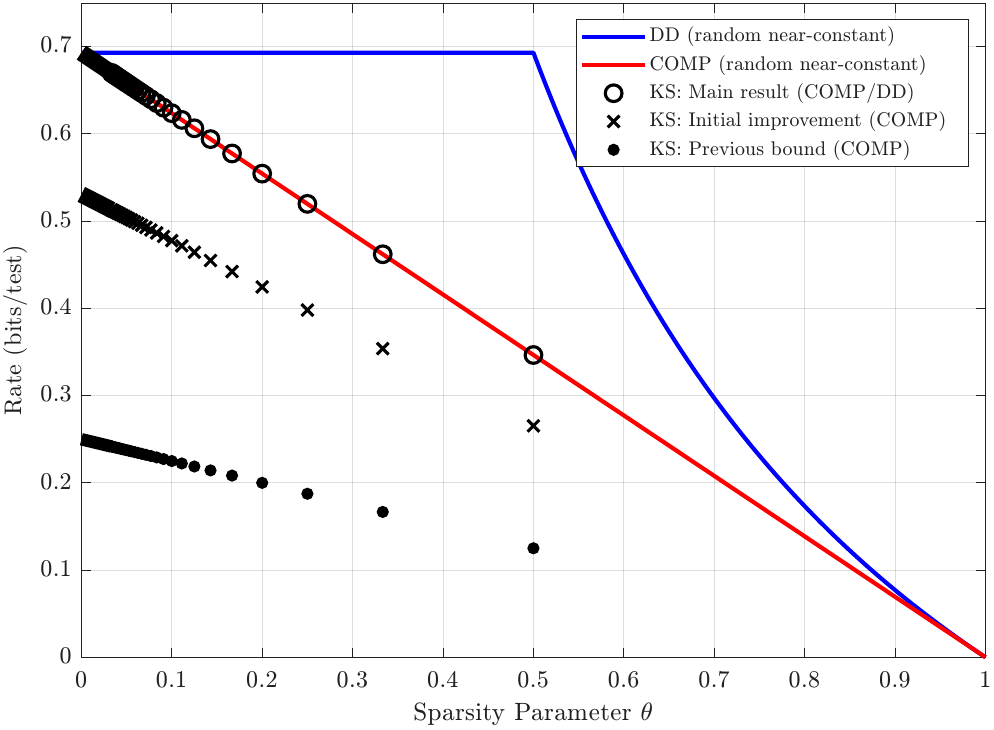}
	\caption{Achievable rates for the Kautz--Singleton construction
		under COMP and DD decoding, compared with random
		near-constant-column-weight designs.
		The Kautz--Singleton results are shown at the admissible sparsity
		exponents $\theta=1/k$.} \vspace*{-3ex}
	\label{fig:KS-rates}
\end{figure}

\subsection{Extension to Thresholds Smaller Than One}

Thus far, the improvement  in Theorem~\ref{thm: warmup} was obtained by replacing the prior work's threshold $2d$ with the more general threshold $(1+\eta)d$, where $\eta>0$. In this section, we push this idea one step further and study whether the same argument can be extended to thresholds of the form $\eta d$ for arbitrary fixed $\eta>0$, including the regime $\eta<1$.

For the finite field $\mathbb F_q$, define
\begin{align}
\mathcal P_{<k}&:=\{f\in\mathbb F_q[x]:\deg f<k\},
\nonumber\\
\mathcal P_{<k}^*&:=\mathcal P_{<k}\setminus\{0\},
\label{eq:polynomial-spaces}
\end{align}
and write
\begin{align}
M:=|\mathcal P_{<k}^*|=q^k-1,
\qquad
M_0:=q^{k-1}-1,
\end{align}
noting that $M_0$ is the number of nonzero polynomials vanishing at any fixed field point. For a polynomial $f$, write
\begin{align}
R(f):=\{a\in\mathbb F_q:f(a)=0\}.
\end{align}
Let $F_1,\ldots,F_d$ be sampled uniformly without replacement from $\mathcal P_{<k}^*$, and define
\begin{align}
U:=\left|\bigcup_{i=1}^dR(F_i)\right|,
\qquad
\mu_d:=\mathbb E U,
\end{align}
i.e., $U$ counts the number of distinct field elements that are roots of at least one of the sampled polynomials.  
We proceed to state a concentration result for $U$, which we then apply to bound the relevant root-covering probability.

\begin{lemma}[Union-size mean and concentration]
\label{lem:betterbound}
For admissible parameters (Definition~\ref{def:admissible}) with $q/d\to c>0$,
\begin{align}
\mu_d
&=q\left(1-\frac{\binom{M-M_0}{d}}{\binom Md}\right),
\label{eq:union-mean}\\
\frac{\mu_d}{d}
&\longrightarrow\rho_c:=c(1-e^{-1/c}).
\label{eq:union-mean-limit}
\end{align}
Moreover, if $d=\Omega(\log^2N)$ we have for any fixed $\eta>0$ that
\begin{align}
\Pr(U-\mu_d\ge\eta d)
=o\left(\frac{1}{N}\right).
\label{eq:union-concentration}
\end{align}
\end{lemma}
\begin{proof}[Proof sketch]
	The exact mean follows by counting the polynomials that vanish at a fixed field element and then computing the probability that none of the sampled polynomials vanish there. To show concentration, we first consider sampling with replacement, in which we show that the normalized union size is self-bounding, thus permitting a standard concentration bound. We then transfer the resulting upper-tail bound to sampling without replacement by a conditioning argument (on the sampled polynomials all being distinct), controlling both the differences in means and the probability of the conditioning event.  This yields a tail bound whose exponent scales as $d/(k-1)=\omega(\log N)$, thus yielding $o(1/N)$ scaling. The full proof is given in Appendix~\ref{app: proofBetterBound}.
\end{proof}

For $1\le m\le M$, let $B_m$ be a uniformly random subset of $\mathcal P_{<k}^*$ of size $m$, and define
\begin{align}
p_m:=\Pr\left(\Psi\subseteq\bigcup_{f\in B_m}R(f)\right),
\label{eq:dd-pm}
\end{align}
where the dependence of $p_m$ on $\Psi$ is left implicit.

\begin{lemma}[Root-covering probability]
\label{lem:root-covering}
Suppose that $q/d\to c>0$ and $d=\Omega(\log^2 N)$ along an admissible sequence (Definition~\ref{def:admissible}), and fix $\beta\in(\rho_c,c)$. Then, for any set of evaluation points $\Psi\subseteq\mathbb F_q$ with $1\le n=|\Psi|\le \beta d$,
\begin{align}
p_d\le\left(\frac{\beta}{c}\right)^n
+o\left(\frac{1}{N}\right).
\label{eq:root-covering-bound}
\end{align}
\end{lemma}
\begin{proof}[Proof sketch]
	By Lemma~\ref{lem:betterbound}, the probability that
	$U$ exceeds $\beta q/c$ is $o(1/N)$.
	Conditional on $U$, the union of roots is uniformly distributed
	over subsets of $\mathbb F_q$ of that size, by
	Lemma~\ref{lem:conditional-covering}.
	Hence, when $U\le\beta q/c$, the probability that all $n$
	evaluation points are covered is at most $(\beta/c)^n$.
	The details are given in Appendix~\ref{app: proofBetterBound}.
\end{proof}

\subsection{Main Bound}
Our main result for COMP decoding is stated as follows.

\begin{theorem}
\label{thm:sharpbound}
For every $\varepsilon>0$, there is a fixed $c>0$ such that, along any admissible sequence (Definition~\ref{def:admissible}) with $q/d\to c$ and $d=\Omega(\log^2N)$, the Kautz--Singleton block length $n$ can be chosen such that COMP recovers a uniformly random defective set of size $d$ with the number of tests satisfying
\begin{align}
T
\le
\left(
\frac{1}{\ln 2}+\varepsilon
\right)
d\log N,
\qquad 1/\ln2\approx1.4427,
\end{align}
while the error probability satisfies $P_e\to0$ as $N \to \infty$.
\end{theorem}
\begin{proof}[Proof sketch]
We follow the proof of Theorem~1 in~\cite{inan2019optimality}, but replace the deterministic root-count threshold by a threshold centered around the actual mean number of covered field elements. For a fixed nondefective item, subtracting its polynomial from the defective polynomials gives a uniform $d$-subset of $\mathcal P_{<k}^*$. The item is covered exactly when the roots of these polynomials contain all tested evaluation points. This event has probability $p_d$ from \eqref{eq:dd-pm}, so a union bound gives
$P_e\le(N-d)p_d$. 
Under the scaling $q/d\to c$, Lemma~\ref{lem:betterbound} gives $\mathbb E U=(\rho_c+o(1))d$. Moreover, fixing $\gamma\in(0,c-\rho_c)$ and letting $\beta=\rho_c+\gamma$, the same lemma implies that $\Pr(U>\beta q/c) = o(1/N)$. 

Conditioning on the complementary event, the conditional-uniformity argument used in~\cite{inan2019optimality} bounds the covering probability by $(\beta/c)^n$. These two steps are summarized in Lemma~\ref{lem:root-covering}, giving $p_d\le(\beta/c)^n+o(1/N)$. For fixed $\delta>0$, choose
\begin{equation}
n=\left\lceil\frac{(1+\delta)\log N}{\log(c/\beta)}\right\rceil.
\label{eq:comp-sketch-n}
\end{equation}
Then $(\beta/c)^n\le N^{-(1+\delta)}$ and $P_e\le N^{-\delta}+o(1)\to0$. Since $T=nq$, taking the fixed slack constants arbitrarily small after the asymptotic limit yields the limiting test coefficient $C(c)=c/\log(c/\rho_c)$. Finally, optimizing over $c>0$ gives $c=1/\ln2$ and $\inf_{c>0}C(c)=1/\ln2$. The full proof, including feasibility of the chosen block length, is given in Appendix~\ref{app: proofSharpBound}.
\end{proof}

\textbf {Comparison with existing works.} The constant obtained in Theorem~\ref{thm:sharpbound} can be compared directly with the best-known bounds for COMP under commonly studied random test designs. For Bernoulli testing, where every item is independently included in every test with an optimized inclusion probability, COMP succeeds with vanishing error probability using
\begin{align}
T
=
\left(
e\ln 2+o(1)
\right)d\log N,
\end{align}
and this threshold is known to be tight for COMP under Bernoulli designs; see \cite{aldridge2014group} and \cite{aldridge2017capacity}. Thus, the corresponding leading constant is $e\ln 2\approx1.884$.

A smaller constant is achievable using random near-constant column-weight designs. In this construction, each item is independently assigned to approximately the same number of tests. It was shown in \cite{aldridge2016improved,johnson2018performance} that COMP exhibits a sharp threshold at
\begin{align}
T
=
\left(
\frac{1}{\ln 2}+o(1)
\right)d\log N.
\end{align}
For deterministic test designs, to the best of our knowledge, the best existing COMP guarantee for explicit designs is given by \cite{inan2019optimality}, summarized in Theorem~\ref{thm: asymBound}. As discussed above, that result yields the asymptotic bound $T\le(4+o(1))d\log N$. Therefore, Theorem~\ref{thm:sharpbound} substantially improves the best known leading constant for deterministic COMP designs, reducing it from $4$ to $1/\ln2\approx1.4427$. Moreover, this matches the leading constant achieved by the best known random design under COMP \cite{johnson2018performance}. (See Figure \ref{fig:KS-rates}.)

\subsection{Infeasibility of Smaller Root Thresholds}
The argument leading to Theorem~\ref{thm:sharpbound} relies on choosing a threshold slightly above $\rho_c d$, and showing that the probability of exceeding this threshold is sufficiently small. It is therefore natural to ask whether the threshold can be reduced further. In this section, we show that $\rho_c d$ represents a fundamental limit for this approach. More precisely, the normalized number of distinct roots $U/d$ converges in probability to $\rho_c$. Consequently, any fixed threshold strictly below $\rho_c d$ is exceeded with probability tending to one. We formalize this through the following lemma.

\begin{lemma}[Infeasibility of smaller thresholds]
\label{lem:alpha-threshold}
Consider any admissible sequence (Definition~\ref{def:admissible}) with $q/d\to c>0$, and suppose that $k^2=o(d)$.\footnote{This condition automatically holds when $d=\Omega(\log^2 N)$, since $k=\frac{\log N}{\log q}$ gives $\frac{k^2}{d}=O\bigl(\frac{1}{\log^2 q}\bigr)=o(1)$.} Then
\begin{align}
\frac{U}{d}
\xrightarrow{\mathrm{P}}
\rho_c,
\end{align}
where $\rho_c = c(1-e^{-1/c})$.  
Hence, for every fixed $\eta<\rho_c$, we have $\Pr\left(U>\eta d\right) \to 1$.
\end{lemma}

\begin{proof}[Proof sketch]
Lemma~\ref{lem:betterbound} gives $\mu_d/d\to\rho_c$. To establish concentration, we reveal the sampled polynomials sequentially and consider the resulting Doob martingale. By coupling two possible choices of the next polynomial through a swap, we show that the corresponding conditional expectations differ by at most $k-1$. Azuma--Hoeffding then gives fluctuations of order $(k-1)\sqrt d=o(d)$, establishing the claimed convergence in probability. The full proof is given in Appendix~\ref{app: proofAlpha}.
\end{proof}

We emphasize that Lemma~\ref{lem:alpha-threshold} only establishes a barrier for the current proof strategy; it is conceivable that a different analysis could overcome this barrier.


\subsection{Extension to DD Decoding}
\label{sec:dd_analysis}

We now turn our attention to DD decoding, as described in Section \ref{sec: Prelim}.  While DD attains a strictly better constant than COMP for random designs when $d = \Theta(N^{\theta})$ \cite{aldridge2014group}, we leave open whether the improved constant can be attained under the Kautz-Singleton construction, and instead only achieve the same constant as COMP.  Such a result is still significant due to the fact that DD guarantees no false positives, whereas COMP guarantees no false negatives, and either of these properties could be preferred depending on the application.

Our analysis is based on the fact that the following conditions are sufficient (albeit not necessary) for DD to succeed: (i) no nondefective item is covered by the defective set, and (ii) each defective item appears as the unique possible defective in at least one positive test. We show that both sources of error can be reduced to the same root-covering event studied in the previous sections. Consequently, the root-covering bound of Lemma~\ref{lem:root-covering} can be applied without changing the leading constant, yielding the following guarantee for DD.
\begin{theorem}
\label{thm:dd-sharpbound}
For every $\varepsilon>0$, there is a fixed $c>0$ such that, along any admissible sequence (Definition~\ref{def:admissible}) with $q/d\to c$ and $d=\Omega(\log^2N)$, the Kautz--Singleton block length $n$ can be chosen so that DD recovers a uniformly random defective set of size $d$ with the number of tests satisfying
\begin{align}
T
\le
\left(
\frac{1}{\ln 2}+\varepsilon
\right)
d\log N,
\end{align}
while the average probability of error satisfies $P_e\to0$. 
\end{theorem}
\begin{proof}[Proof sketch]
For a defective set $S$, let $C(S)$ denote the set of nondefective items that remain undetected after the first stage of DD.  The following conditions suffice for the DD estimate to be correct: (i) $C(S)=\varnothing$, and (ii) every defective item has at least one tested evaluation coordinate at which it differs from all other defectives.

Both failure events can be expressed in terms of the same root-covering probability. For a fixed nondefective item, translation by its corresponding polynomial shows that the event of being covered is the event that $d$ uniformly sampled distinct nonzero Reed--Solomon polynomials have roots covering all evaluation points. This is the probability $p_d$ defined in \eqref{eq:dd-pm}.  By similar reasoning, for a fixed defective item, the absence of a ``private'' evaluation coordinate (i.e., one not shared with any other defective) is governed by $p_{d-1}$. A uniform-subset deletion coupling gives $p_{d-1}\le p_d$. Consequently,
\begin{equation}
P_e\le(N-d)p_d+dp_{d-1}\le Np_d.
\end{equation}
Applying Lemma~\ref{lem:root-covering}, we obtain
$
p_d\le(\beta/c)^n+o(1/N),
$
for any fixed $\beta\in(\rho_c,c)$. Fixing $\delta>0$ and choosing $n$ as in \eqref{eq:comp-sketch-n} makes the first term at most $N^{-(1+\delta)}$, yielding $P_e\le N^{-\delta}+o(1)\to0$. Taking $\beta\downarrow\rho_c$ and $\delta\downarrow0$ gives the limiting test coefficient $c/\log(c/\rho_c)$, whose minimum over $c>0$ is $1/\ln2$. The full proof is given in Appendix~\ref{app: proofDD}.
\end{proof}

\textbf{Comparison with existing works.} For random test designs, the performance of DD has been studied extensively under both Bernoulli and near-constant tests-per-item constructions. In the regime $d=\Theta(N^\theta)$, \cite{aldridge2014group} showed that DD with a Bernoulli design succeeds using
$T=\left(
e\ln 2\,\max\{\theta,1-\theta\}+o(1)
\right)
d\log N$.
In particular, in the sparse limit $\theta\to0$, this gives the leading constant $e\ln2\approx1.884$. This was subsequently improved in \cite{johnson2018performance} using the random near-constant tests-per-item design, attaining
\begin{align}
T
=
\left(
\frac{\max\{\theta,1-\theta\}}{\ln2}
+o(1)
\right)
d\log N.
\end{align}
In particular, in the sparse limit $\theta\to0$, the corresponding leading constant is $1/\ln2\approx1.4427$. Theorem~\ref{thm:dd-sharpbound} therefore matches the near-constant tests-per-item constant in the regime $d=N^{o(1)}$. For the admissible fixed exponents $\theta=1/k$ (with integer $k \ge 2$), a gap remains to the smaller random-design DD constant; our guarantee is only at the COMP-level threshold.  See Figure \ref{fig:KS-rates} for a summary.




\section{Conclusion}

In this work, we revisited the probabilistic analysis of the Kautz--Singleton construction, improving the constant in the number of tests for COMP decoding and deriving the same constant for DD decoding.   For COMP, our result matches the best-known random near-constant-column-weight constant, while for DD it matches the corresponding random-design constant in the sparse regime $d = N^{o(1)}$.  

We conclude with some possible questions for future work:
\begin{itemize}
	\item Can we handle the regime $d = \Theta(N^\theta)$ for general $\theta \in (0,1)$ instead of only $\theta = 1/k$ (for integer-valued $k \ge 2$)?
	\item Can the Kautz--Singleton construction attain the improved DD threshold attained by randomized designs?
	\item How many tests does an information-theoretically optimal decoder need under the  Kautz--Singleton construction?
\end{itemize}

\newpage
\bibliographystyle{IEEEtran}
\bibliography{ISIT}

\appendices

\clearpage

\section*{Table of Notation}

\begin{table}[h]
\renewcommand*{\arraystretch}{1.25}
\begin{tabularx}{\columnwidth}{p{0.29\columnwidth}X}
\toprule
Notation & Meaning \\
\midrule

$N$ 
& Number of items \\

$d$ 
& Number of defective items \\

$S\subseteq[N]$ 
& Uniformly random defective set with $|S|=d$ \\

$q$ 
& Size of the finite field $\mathbb F_q$; $q=\Theta(d)$ \\

$c_N,\ c$
& Ratio $q/d$ and its positive limiting value \\

$k$ 
& Dimension of the underlying Reed--Solomon code, with $N=q^k$ \\

$n$ 
& Number of Reed--Solomon evaluation points \\

$T=nq$ 
& Total number of binary group tests \\

$\Psi=\{\alpha_1,\ldots,\alpha_n\}$ 
& Set of Reed--Solomon evaluation points in $\mathbb F_q$ \\

$\mathcal P_{<k}$ 
& Set of all polynomials in $\mathbb F_q[x]$ of degree less than $k$ \\

$\mathcal P_{<k}^*$ 
& $\mathcal P_{<k}\setminus\{0\}$ \\

$R(f)$ 
& Root set $\{a\in\mathbb F_q:f(a)=0\}$ of a polynomial $f$ \\

$U$ 
& Number of distinct field elements covered by the roots of the sampled polynomials \\

$\rho_c$ 
& $c(1-e^{-1/c})$, the asymptotic normalized mean of $U$ \\

$P_e$ 
& Average probability of decoding error \\

$A,\ A_i$
& Binary test matrix in $\{0,1\}^{T\times N}$ and its $i$-th column \\

$M=q^k-1$
& Number of polynomials in $\mathcal P_{<k}^*$ \\

$M_0=q^{k-1}-1$
& Number of polynomials in $\mathcal P_{<k}^*$ vanishing at a fixed point \\

$\mu_d,\ \widetilde\mu_d$
& Means of $U$ and $\widetilde U$, respectively \\

$C(S)$ 
& Set of nondefective items covered by the defective columns \\

$P_{f_i}^{\Psi}(S\setminus\{i\})$ 
& Private evaluation points of defective polynomial $f_i$ \\

$p_m$ 
& Probability that the roots of $m$ uniformly sampled distinct nonzero polynomials cover all points in $\Psi$ \\
$\xrightarrow{\mathrm{P}}$
& Convergence in probability \\

$O(\cdot),\,o(\cdot)$
& Standard asymptotic order notation \\

$\gtrsim$
& Inequality up to a positive constant factor asymptotically \\

$[N]$
& Set $\{1,\ldots,N\}$ \\

$\operatorname{supp}(A_i)$
& Support of the $i$-th column of the test matrix \\

$\mathbf{1}\{\cdot\}$
& Indicator function \\

$\binom{\mathcal A}{m}$
& Collection of all $m$-element subsets of the set $\mathcal A$ \\

$\operatorname{Unif}(\mathcal A)$
& Uniform distribution over the finite set $\mathcal A$ \\

$(M)_d$
& Falling factorial $M(M-1)\cdots(M-d+1)$ \\

$F_1,\ldots,F_d$
& Polynomials sampled uniformly without replacement from $\mathcal P_{<k}^*$ \\

$\widetilde F_1,\ldots,\widetilde F_d$
& Polynomials sampled independently with replacement from $\mathcal P_{<k}^*$ \\

$\widetilde U$
& Number of distinct roots covered in the with-replacement model \\
\bottomrule
\end{tabularx}
\label{tab:notation}
\end{table}

\newpage

%

\section{Proof of Theorem~\ref{thm: warmup} (Initial Refinement)}\label{app: proofthmwarmup}

We first state a useful lemma that will be used below and re-used in later appendices.

\begin{lemma}[Conditional uniformity and covering]
	\label{lem:conditional-covering}
	Let $F_1,\ldots,F_d$ be sampled uniformly without replacement from $\mathcal P_{<k}^*$, and let $h=\prod_{j=1}^d F_j$. Then, we have the following:\footnote{All conditional statements are implicitly restricted to conditioning events of positive probability.}
	
	\medskip
	\noindent\textup{(i)} For any integer $\ell\in[0,q]$, conditional on $|R(h)|=\ell$, the set $R(h)$ is uniform over all $\ell$-subsets of $\mathbb F_q$.
	
	\medskip
	\noindent\textup{(ii)} For any fixed $\Psi\subseteq\mathbb F_q$ with $1\le n=|\Psi|\le q$ and any integer $\ell\in[n,q]$,
	\begin{gather}
		\Pr(\Psi\subseteq R(h)\mid |R(h)|=\ell)
		=\frac{\binom{q-n}{\ell-n}}{\binom q\ell},
		\label{eq:sharp-conditional}\\
		\frac{\binom{q-n}{\ell-n}}{\binom q\ell}
		=\prod_{a=0}^{n-1}\frac{\ell-a}{q-a},
		\label{eq:ratio-product}
	\end{gather}
	and for $\ell<n$, the conditional probability is zero. Consequently, for real $L$ with $n\le L\le q$, we have
	\begin{align}
		&\Pr(\Psi\subseteq R(h)\mid |R(h)|\le L)
		\nonumber\\
		&\le\prod_{a=0}^{n-1}\frac{\lfloor L\rfloor-a}{q-a}
		\le\left(\frac{L}{q}\right)^n.
		\label{eq:conditional-cutoff}
	\end{align}
	The same upper bound holds for the joint event
	$\{\Psi\subseteq R(h),\,|R(h)|\le L\}$, and if $0\le L<n$ then the joint probability is zero.
\end{lemma}

\begin{proof}
Part (i) is precisely Claim~1 of~\cite{inan2019optimality}. For part (ii), there are $\binom q\ell$ possible root sets of size $\ell$. Exactly $\binom{q-n}{\ell-n}$ contain $\Psi$, since their remaining $\ell-n$ elements must be chosen from $\mathbb F_q\setminus\Psi$. This proves \eqref{eq:sharp-conditional}, and cancellation of factorials gives \eqref{eq:ratio-product}. The statement on $\ell < n$ trivially holds since a set of size less than $n$ cannot contain $\Psi$.

Conditioning on the possible values $\ell\le\lfloor L\rfloor$ of $|R(h)|$, the probability in \eqref{eq:conditional-cutoff} is a weighted average of these covering probabilities. For $\ell\ge n$, each factor in \eqref{eq:ratio-product} is nondecreasing in $\ell$, while values $\ell<n$ contribute zero. Hence, the average is at most the product evaluated at $\ell=\lfloor L\rfloor$. For $0\le a\le n-1$,
\begin{align}
\frac{\lfloor L\rfloor-a}{q-a}
\le\frac{\lfloor L\rfloor}{q}
\le\frac{L}{q},
\end{align}
where the first inequality uses $\lfloor L\rfloor\le q$. Multiplying terms gives the last bound in \eqref{eq:conditional-cutoff}. Further multiplication by $\Pr(|R(h)|\le L)\le1$ establishes the joint-event claim.
\end{proof}

\begin{proof}[Proof of Theorem~\ref{thm: warmup}]
Fix $\varepsilon>0$. We follow the proof of Theorem~1 in~\cite{inan2019optimality}, with two modifications: we replace the root threshold $2d$ by $(1+\eta)d$ for some $\eta>0$, and we optimize the choice of the Reed--Solomon alphabet size $q$.

Recall the Kautz--Singleton matrix $A$, with $N=q^k$ items and $T=nq$ tests, and its set of evaluation points $\Psi$ from Section~\ref{sec: Prelim}.

For a set $s\subseteq[N]$ with $|s|=d$, let $\mathcal E^s$ denote the event that at least one column indexed by $[N]\setminus s$ is covered by the columns indexed by $s$. For $i\notin s$, let $\mathcal E_i^s$ denote the event that the column $A_i$ is covered by the columns indexed by $s$. Since COMP makes an error only if at least one nondefective column is covered by the defective columns, a union bound gives
\begin{align}
P_e
&\le
\frac{1}{\binom{N}{d}}
\sum_{\substack{s\subseteq[N]\\|s|=d}}
\sum_{i\in[N]\setminus s}
\mathbf{1}\{\mathcal E_i^s\} \nonumber\\
&=
\frac{N-d}{N}
\sum_{i\in[N]}
\Pr\left(\mathcal E_i^{S_{[N]\setminus\{i\}}}\right),
\label{eq:warmup_union}
\end{align}
where $S_{[N]\setminus\{i\}}$ is uniformly distributed over all subsets of $[N]\setminus\{i\}$ of cardinality $d$, and the term $\frac{N-d}{N}$ arises as the fraction of $d$-sized subsets that exclude a given item $i$, i.e., $\frac{\binom{N-1}{d}}{\binom Nd}=\frac{N-d}{N}$.

Fix $i\in[N]$. For each column $A_j$, let $f_j(X)$ denote its corresponding polynomial of degree at most $k-1$. We have
\begin{align}
&\Pr\left(\mathcal E_i^{S_{[N]\setminus\{i\}}}\right)\\
&=
\Pr\left(
f_i(\alpha)
\in
\{f_j(\alpha):j\in S_{[N]\setminus\{i\}}\}
\text{ for all }\alpha\in\Psi
\right)
\nonumber\\
&=
\Pr\left(
0
\in
\{f_j(\alpha)-f_i(\alpha):
j\in S_{[N]\setminus\{i\}}\}
\text{ for all }\alpha\in\Psi
\right).
\label{eq:translated-cover}
\end{align}

The Reed--Solomon code contains all polynomials over $\mathbb{F}_q$ of degree at most $k-1$. Hence, after translating every polynomial by $-f_i$, the collection
\begin{equation}\label{eq: transset}
\left\{
f_j(X)-f_i(X):j\in[N]\setminus\{i\}
\right\}
\end{equation}
is precisely the collection of all polynomials in $\mathcal P_{<k}^*$. Therefore, the probability in~\eqref{eq:translated-cover} is equal to the probability obtained by selecting uniformly without replacement $d$ polynomials $F_1,\ldots,F_d$ from $\mathcal P_{<k}^*$ and requiring that every element of $\Psi$ be a root of at least one selected polynomial.

Define
\begin{equation}
    h(X)=\prod_{j=1}^d F_j(X)
\end{equation}
so that $R(h)=\bigcup_{j=1}^dR(F_j)$. Then, the equivalence noted following \eqref{eq: transset} gives
\begin{equation}
\mathcal E_i^{S_{[N]\setminus\{i\}}}
\quad\Longleftrightarrow\quad
R(h)\supseteq\Psi.
\label{eq:cover-root-equivalence}
\end{equation}

We first control the cardinality of $R(h)$. Let $r$ denote the number of distinct roots of a uniformly random nonzero polynomial over $\mathbb{F}_q$ of degree at most $k-1$. The following lemma summarizes a result established in Appendix~A of~\cite{inan2019optimality} (shortly after Lemma 1 therein).

\begin{lemma}[Appendix A of \cite{inan2019optimality}]
\label{lem:root-moments}
Let $q$ be a prime power and $k\ge 1$. Let $f(X)$ be a uniformly random nonzero polynomial over $\mathbb{F}_q$ of degree at most $k-1$, and let $r$ denote the number of distinct roots of $f(X)$ in $\mathbb{F}_q$. Then,
\begin{equation}
    \mathbb{E}[r]\le 1
    \qquad\text{and}\qquad
    \mathbb{E}[r^2]<6.
    \label{eq:root-moments}
\end{equation}
\end{lemma}
This lemma implies that
\begin{equation}
    \operatorname{Var}(r)<6.
\end{equation}
Furthermore, every nonzero polynomial of degree at most $k-1$ has at most $k-1$ roots, so
\begin{equation}
    0\le r\le k-1<k.
\end{equation}

For each $j\in[d]$, let $r_j:=|R(F_j)|$ be its root count. Every root of $h$ is a root of at least one of its factors, and therefore
\begin{equation}
    |R(h)|
    \le
    \sum_{j=1}^dr_j.
\label{eq:root-union}
\end{equation}

Let $\eta>0$ be fixed. Using~\eqref{eq:root-union} and $\mathbb{E}[r_j]\le1$, we obtain
\begin{align}
\Pr\left(|R(h)|>(1+\eta)d\right)
&\le
\Pr\left(
\sum_{j=1}^dr_j>(1+\eta)d
\right)
\nonumber\\
&=
\Pr\left(
\frac{1}{d}\sum_{j=1}^dr_j>1+\eta
\right)
\nonumber\\
&\le
\Pr\left(
\frac{1}{d}
\sum_{j=1}^d
\left(r_j-\mathbb{E}[r_j]\right)
>\eta
\right).
\label{eq:root-tail-start}
\end{align}

The variables $\{r_j:j\in[d]\}$ form a sample drawn uniformly without replacement from the finite population consisting of the root counts of all nonzero polynomials of degree at most $k-1$. Applying Bernstein's inequality for sampling without replacement, together with~\eqref{eq:root-moments}, gives
\begin{align}
\Pr\left(|R(h)|>(1+\eta)d\right)
&\le
\exp\left(
-\frac{d\eta^2}
{2\operatorname{Var}(r)+\frac{2}{3}k\eta}
\right)
\nonumber\\
&\le
\exp\left(
-\frac{d\eta^2}
{12+\frac{2}{3}k\eta}
\right).
\label{eq:root-tail}
\end{align}
Since $N=q^k$, we have
\begin{equation}
    k=\frac{\log N}{\log q}.
\end{equation}
Hence, under the assumed regime $d=\Omega(\log^2 N)$, the exponent in~\eqref{eq:root-tail} scales as
\begin{equation}
    \Omega\Big(\frac{d}{k}\Big)
    =
    \Omega(\log N\log q).
\end{equation}
Therefore, for every fixed $\eta>0$, there exists a constant $c_\eta>0$ such that, for all sufficiently large $N$,
\begin{equation}
    \Pr\left(|R(h)|>(1+\eta)d\right)
    \le
    N^{-c_\eta\log q}.
    \label{eq:root-tail-final}
\end{equation}

Next, choose a fixed limit $c>1+\eta$ and take $c_N=q/d\to c$ (see Definition \ref{def:admissible}). Then $c_N>1+\eta$ for all sufficiently large $N$. The choice of $n$ in \eqref{eq:n-choice} below has $n=\Theta(\log N)$, so $d=\Omega(\log^2N)$ gives $n=o(d)$ and hence $n\le\lfloor(1+\eta)d\rfloor$ eventually. Applying Lemma~\ref{lem:conditional-covering}(ii) with $L=(1+\eta)d<q$ then gives
\begin{align}
&\Pr\left(R(h)\supseteq\Psi\,\middle|\,|R(h)|\le(1+\eta)d\right)
\nonumber\\
&\le\left(\frac{(1+\eta)d}{q}\right)^n
=\left(\frac{1+\eta}{c_N}\right)^n.
\label{eq:conditional-cover-final}
\end{align}
Note that the conditioning event has positive probability for all sufficiently large $N$ by \eqref{eq:root-tail-final}.

Combining~\eqref{eq:cover-root-equivalence},
\eqref{eq:root-tail-final}, and
\eqref{eq:conditional-cover-final}, we obtain for every $i\in[N]$ that
\begin{align}
\Pr\left(\mathcal E_i^{S_{[N]\setminus\{i\}}}\right)
&\le
\Pr\left(
R(h)\supseteq\Psi
\,\middle|\,
|R(h)|\le(1+\eta)d
\right)
\nonumber\\
&\qquad+
\Pr\left(
|R(h)|>(1+\eta)d
\right)
\nonumber\\
&\le
\left(\frac{1+\eta}{c_N}\right)^n
+
N^{-c_\eta\log q}.
\label{eq:single-column-final}
\end{align}

Substituting~\eqref{eq:single-column-final} into~\eqref{eq:warmup_union} and using $(N-d)/N\le1$ yields
\begin{align}
P_e
&\le
N\left(\frac{1+\eta}{c_N}\right)^n
+
N^{1-c_\eta\log q}.
\label{eq:Pe-before-n}
\end{align}
Since we assume that $q\to\infty$ (see Definition \ref{def:admissible}), we have
\begin{equation}
    N^{1-c_\eta\log q}
    =
    N^{-\Omega_\eta(\log q)},
\end{equation}
where $\Omega_\eta(\cdot)$ represents $\Omega(\cdot)$ notation while highlighting that the hidden constant may depend on $\eta$.

Let $\delta>0$, and choose
\begin{equation}
n
=
\left\lceil
\frac{(1+\delta)\log N}
{\log\left(c_N/(1+\eta)\right)}
\right\rceil.
\label{eq:n-choice}
\end{equation}
Because $c_N\to c>1+\eta$, the denominator converges to a positive constant and $n=\Theta(\log N)$, as used above. Note also that $q=\Theta(d)$ and $d=\Omega(\log^2N)$ give $n=o(q)$, while $k=\log N/\log q$ gives $n/k=\Theta(\log q)\to\infty$, so that $k\le n\le q$ eventually. 

We further bound the first term in \eqref{eq:Pe-before-n} as follows:
\begin{align}
N\left(\frac{1+\eta}{c_N}\right)^n
&\le
N
\left(\frac{1+\eta}{c_N}\right)^{
\frac{(1+\delta)\log N}
{\log(c_N/(1+\eta))}
}
\nonumber\\
&=
N\,2^{-(1+\delta)\log N}
\nonumber\\
&=
N^{-\delta}.
\end{align}
Therefore,
\begin{equation}
    P_e
    \le
    N^{-\delta}
    +
    N^{-\Omega_\eta(\log q)},
\label{eq:warmup-Pe}
\end{equation}
and in particular $P_e\to0$ as $N\to\infty$.

It remains to optimize the number of tests. From $T=nq$, $q=c_Nd$, $c_N\to c$, and~\eqref{eq:n-choice}, we have
\begin{equation}
T
=
\left(
\frac{c(1+\delta)}
{\log\left(c/(1+\eta)\right)}
+o(1)
\right)d\log N.
\label{eq:test-general}
\end{equation}
For fixed $\eta>0$, write
\begin{equation}
    x=\frac{c}{1+\eta}>1.
\end{equation}
Then, the leading constant in~\eqref{eq:test-general} is
\begin{equation}
    (1+\delta)(1+\eta)\frac{x}{\log x}.
\end{equation}
Since $\log$ is base $2$, we have
\begin{equation}
    \frac{x}{\log x}
    =
    (\ln 2)\frac{x}{\ln x}.
\end{equation}
The function $x/\ln x$ is minimized over $x>1$ at $x=e$, because
\begin{equation}
\frac{\mathrm{d}}{\mathrm{d}x}
\left(\frac{x}{\ln x}\right)
=
\frac{\ln x-1}{(\ln x)^2},
\end{equation}
which vanishes uniquely at $x=e$. Hence, the optimal choice is
\begin{equation}
    c=e(1+\eta),
\end{equation}
and~\eqref{eq:test-general} becomes
\begin{equation}
T
=
\left(
(1+\delta)(1+\eta)e\ln 2
+o(1)
\right)d\log N.
\label{eq:test-optimized}
\end{equation}

Finally, since $\varepsilon>0$ is arbitrary, we may choose fixed $\delta>0$ and $\eta>0$ sufficiently small so that
\begin{equation}
    (1+\delta)(1+\eta)e\ln 2
    <
    e\ln 2+\frac{\varepsilon}{2}.
\end{equation}
We choose admissible parameters with $q/d\to c=e(1+\eta)$, as permitted by Definition~\ref{def:admissible}. The $o(1)$ term is smaller than $\varepsilon/2$ for all sufficiently large $N$. Therefore,
\begin{equation}
    T
    \le
    (e\ln 2+\varepsilon)d\log N
\end{equation}
for all sufficiently large $N$, while~\eqref{eq:warmup-Pe} ensures that $P_e\to0$. This completes the proof.
\end{proof}

\section{Proofs of Lemmas~\ref{lem:betterbound} and~\ref{lem:root-covering} (Concentration and Root Covering Probability)}\label{app: proofBetterBound}
\subsection{Proof of Lemma~\ref{lem:betterbound}}\label{app: union-concentration}
Recall that $M = q^k - 1$ represents the size of $\mathcal P_{<k} \setminus \{0\}$.  The definition $M_0=q^{k-1}-1$ is similarly justified as follows:  
For any fixed $a \in \mathbb{F}_q$, the condition $f(a)=0$
imposes one nonzero linear constraint on the $k$ coefficients of
$f \in \mathcal P_{<k}$. Hence, exactly $q^{k-1}$
polynomials satisfy this condition, including when $a=0$. The zero
polynomial is one of them, so excluding it leaves precisely
$q^{k-1}-1$ such polynomials. 

By the preceding discussion, under sampling without replacement, we have
\begin{align}
\Pr\left(
F_j(a)\neq0
\text{ for every }j\in[d]
\right)
=
\frac{\binom{M-M_0}{d}}
{\binom{M}{d}}.
\label{eq:single-point-avoidance}
\end{align}
Linearity of expectation then gives
\begin{align}
\mu_d
=
q\left[
1-
\frac{\binom{M-M_0}{d}}
{\binom{M}{d}}
\right].
\label{eq:sharp-exact-mu}
\end{align}

We next determine the asymptotic behavior of $\mu_d$ under $q/d\to c$. Observe that
\begin{align}
\frac{\binom{M-M_0}{d}}
{\binom{M}{d}}
=
\prod_{j=0}^{d-1}
\left(
1-\frac{M_0}{M-j}
\right).
\label{eq:avoidance-product}
\end{align}
Since $k\ge2$ and $q=\Theta(d)$, we have uniformly over $0\le j\le d-1$ that
\begin{align}
\frac{M_0}{M-j}
=
\frac{1}{q}
+
O\left(
\frac{1}{q^2}
\right),
\label{eq:sharp-ratio-expansion}
\end{align}
where the expansion follows from $M=q^k-1$, $M_0=q^{k-1}-1$, and $j=O(q)$ uniformly over the stated range.  Using $\ln(1-x)=-x+O(x^2)$ then gives
\begin{align}
\ln
\frac{\binom{M-M_0}{d}}
{\binom{M}{d}}
&=
\sum_{j=0}^{d-1}
\ln\left(
1-\frac{M_0}{M-j}
\right)
\nonumber\\
&=
-\frac{d}{q}
+
O\left(
\frac{d}{q^2}
\right)
\nonumber\\
&=
-\frac{1}{c}
+
o(1).
\end{align}
Therefore,
\begin{align}
\frac{\binom{M-M_0}{d}}
{\binom{M}{d}}
\longrightarrow
e^{-1/c}. \label{eq:mu_and_alpha}
\end{align}
Substituting this into \eqref{eq:sharp-exact-mu} gives
\begin{align}
\frac{\mu_d}{d}
\longrightarrow
\rho_c.
\label{eq:sharp-alpha}
\end{align}

We now compare sampling without replacement to independent sampling. 
Let $\widetilde F_1,\ldots,\widetilde F_d$ be sampled independently and uniformly from $\mathcal P_{<k}^*$, and define
\begin{align}
\widetilde U
:=
\left|
\bigcup_{i=1}^d R(\widetilde F_i)
\right|.
\end{align}
Let
\begin{align}
\widetilde\mu_d
:=
\mathbb{E}\widetilde U.
\end{align}

\begin{lemma}[With-replacement concentration]
\label{lem:wor-concentration}
For admissible parameters with $q/d\to c>0$, the following statements hold:

\medskip
\noindent\textup{(i)} The with-replacement mean satisfies
\begin{gather}
\widetilde\mu_d
=q\left[1-\left(1-\frac{M_0}{M}\right)^d\right],
\label{eq:tilde-mu}\\
\frac{\widetilde\mu_d}{d}\longrightarrow\rho_c,
\label{eq:alpha-c}
\end{gather}
where $\rho_c = c(1-e^{-1/c})$.  

\medskip
\noindent\textup{(ii)} If $d\le M/2$, then for every $t\ge0$,
\begin{align}
&\Pr(U-\mu_d\ge t)
\nonumber\\
&\le\exp\left(
\frac{d(d-1)}{M}
-\frac{t^2}{2(k-1)\widetilde\mu_d+\frac{2}{3}(k-1)t}
\right).
\label{eq:wor-concentration}
\end{align}
\end{lemma}

\begin{proof}
For part (i), the counting argument at the start of this appendix gives probability $M_0/M$ that one sampled polynomial vanishes at any fixed $a\in\mathbb F_q$.  Hence, by independence, the probability that none of the $d$ polynomials exhibit such vanishing behavior is $(1-M_0/M)^d$. Summing the complementary probabilities over $a\in\mathbb F_q$ proves \eqref{eq:tilde-mu}. Moreover, by \eqref{eq:sharp-ratio-expansion} (with $j=0$), we have $dM_0/M\to1/c$ and $d(M_0/M)^2\to0$, giving
\begin{align}
d\ln\left(1-\frac{M_0}{M}\right)
=-\frac{dM_0}{M}+O\left(d\left(\frac{M_0}{M}\right)^2\right)
\longrightarrow-\frac1c.
\end{align}
Together with $q/d\to c$, this gives \eqref{eq:alpha-c}.

For part (ii), we first bound the with-replacement upper tail.
For each $i\in[d]$, define
\begin{align}
\widetilde U_i
:=
\left|
\bigcup_{j\neq i}R(\widetilde F_j)
\right|.
\end{align}
Each sampled polynomial is nonzero and has at most $k-1$ roots, so
\begin{align}
0
\le
\widetilde U-\widetilde U_i
\le
k-1.
\end{align}
Moreover, $\widetilde U-\widetilde U_i$ counts the roots covered by $\widetilde F_i$ and by no other sampled polynomial. Since each root can be uniquely covered by at most one polynomial,
\begin{align}
\sum_{i=1}^d
\left(
\widetilde U-\widetilde U_i
\right)
\le
\widetilde U.
\end{align}
Consequently, $Z=\widetilde U/(k-1)$ is self-bounding.\footnote{A nonnegative function $Z$ of independent inputs is self-bounding if, for each $i$, there is a function $Z_i$ not depending on the $i$th input such that $0\le Z-Z_i\le1$ and $\sum_i(Z-Z_i)\le Z$. Here we take $Z_i=\widetilde U_i/(k-1)$.}
Applying the Bernstein-type concentration inequality for
self-bounding functions \cite{boucheron2000sharp}, we obtain,
for every $s\ge 0$,
\begin{align}
\Pr
\left(
\widetilde U-\widetilde\mu_d\ge s
\right)
\le
\exp\left(
-\frac{s^2}
{2(k-1)\widetilde\mu_d+\frac{2}{3}(k-1)s}
\right).
\label{eq:self-bound-wr}
\end{align}

We next compare the means. For each $a\in\mathbb F_q$, \eqref{eq:single-point-avoidance} and \eqref{eq:avoidance-product} give
\begin{align}
&\Pr\left(F_i(a)\neq0\text{ for every }i\in[d]\right)
\nonumber\\
&\quad=\prod_{j=0}^{d-1}\left(1-\frac{M_0}{M-j}\right)
\nonumber\\
&\quad\le\left(1-\frac{M_0}{M}\right)^d
\nonumber\\
&\quad=\Pr\left(\widetilde F_i(a)\neq0\text{ for every }i\in[d]\right),
\label{eq:sampling_line3}
\end{align}
where inequality holds factorwise because $M-j\le M$; note that the factors are nonnegative since $d\le M/2$ and $M_0/M<1/2$. Taking complements and summing over $a\in\mathbb F_q$ yields
\begin{align}
\mu_d\ge\widetilde\mu_d.
\label{eq:mu-comparison}
\end{align}

Let
\begin{align}
\mathcal D
:=
\left\{
\widetilde F_1,\ldots,\widetilde F_d
\text{ are pairwise distinct}
\right\}.
\end{align}
Conditioned on $\mathcal D$, the ordered tuple $(\widetilde F_1,\ldots,\widetilde F_d)$ is distributed as an ordered uniform sample without replacement from $\mathcal P_{<k}^*$. Moreover,
\begin{align}
\Pr(\mathcal D)
=
\frac{(M)_d}{M^d},
\end{align}
where $(M)_d:=M(M-1)\cdots(M-d+1)$.

Using the conditional representation of the without-replacement model, we have for every $t\ge0$ that
\begin{align}
\Pr(U-\mu_d\ge t)
&=
\Pr
\left(
\widetilde U\ge\mu_d+t
\,\middle|\,
\mathcal D
\right)
\\
&\le
\frac{1}{\Pr(\mathcal D)}
\Pr
\left(
\widetilde U-\widetilde\mu_d
\ge
\mu_d+t-\widetilde\mu_d
\right).
\end{align}
By \eqref{eq:mu-comparison}, $\mu_d+t-\widetilde\mu_d\ge t$. Applying \eqref{eq:self-bound-wr} therefore yields
\begin{align}
\Pr(U-\mu_d\ge t)
\le
\frac{M^d}{(M)_d}
\exp\left(
-\frac{t^2}
{2(k-1)\widetilde\mu_d+\frac{2}{3}(k-1)t}
\right).
\label{eq:wor-precollision}
\end{align}

Since $d\le M/2$, we have $j/M\le1/2$ for $0\le j\le d-1$. Using $-\ln(1-x)\le2x$ on $[0,1/2]$ gives
\begin{align}
\ln\left(
\frac{M^d}{(M)_d}
\right)
&=
\sum_{j=0}^{d-1}
-\ln\left(
1-\frac{j}{M}
\right)
\\
&\le
2\sum_{j=0}^{d-1}\frac{j}{M}
\\
&=
\frac{d(d-1)}{M}.
\end{align}
Substituting this into \eqref{eq:wor-precollision} proves part (ii).
\end{proof}

We now continue the proof of Lemma~\ref{lem:betterbound}. 
Since $N=q^k$ and $k\ge2$, we have
\begin{align}
N\ge q^2=c_N^2d^2.
\label{eq:N_lower}
\end{align}
Combining this with $M=N-1$ and $c_N\to c>0$ gives $d/M=O(1/d)\to0$, so the condition $d\le M/2$ in Lemma~\ref{lem:wor-concentration}(ii) holds eventually. We set $t=\eta d$ in that bound and use part (i) to obtain
\begin{align}
&\Pr
\left(
U-\mathbb{E}U
\ge
\eta d
\right)
\nonumber\\
&\le
\exp\left(
\frac{d(d-1)}{N-1}
-
\frac{\eta^2d^2}
{2(k-1)\widetilde\mu_d
+\frac{2}{3}(k-1)\eta d}
\right)
\nonumber\\
&=
\exp\left[
\frac{d(d-1)}{N-1}
-
\left(
\frac{\eta^2}
{2\rho_c+\frac{2}{3}\eta}
+o(1)
\right)
\frac{d}{k-1}
\right].
\label{eq:main-centered-bound}
\end{align}

We proceed to bound the terms in the exponent.  
Using $N=q^k$, we have
\begin{align}
k-1
=
\frac{\log(N/q)}{\log q},
\end{align}
and therefore
\begin{align}
\frac{d}{k-1}
=
\frac{d\log q}{\log(N/q)}.
\label{eq:d-over-k}
\end{align}
Since $d = \Omega( \log^2N )$ and $\log(N/q)\le\log N$, this simplifies to
\begin{align}
\frac{d}{k-1} = \Omega\big(\log N\log q\big).
\end{align}
Again using the fact that $q \to \infty$, we conclude that
\begin{align}
\frac{d}{k-1}
=
\omega(\log N).
\label{eq:super-log}
\end{align}

On the other hand, since $N\ge c_N^2d^2$ and $c_N\to c>0$ (see \eqref{eq:N_lower}), we have
\begin{align}
\frac{d(d-1)}{N-1}
=
O(1).
\end{align}
Combining these findings with \eqref{eq:super-log}, and noting that $\frac{\eta^2}{2\rho_c+\frac{2}{3}\eta}$ is a positive constant, 
we obtain
\begin{align}
&\left(
\frac{\eta^2}
{2\rho_c+\frac{2}{3}\eta}
+o(1)
\right)
\frac{d}{k-1}
\nonumber\\
&\qquad
-
\ln N
-
\frac{d(d-1)}{N-1}
\longrightarrow+\infty.
\label{eq:log_simplification}
\end{align}
Applying this to \eqref{eq:main-centered-bound} gives
\begin{align}
\Pr
\left(
U-\mathbb{E}U
\ge
\eta d
\right)
=
o\left(
\frac{1}{N}
\right).
\end{align}

\subsection{Proof of Lemma~\ref{lem:root-covering}}\label{app: root-covering}

Fix $\beta\in(\rho_c,c)$ and set $\gamma=(\beta-\rho_c)/2>0$. By Lemma~\ref{lem:betterbound} and $q/d\to c$, we have
\begin{align}
\frac{\beta q}{c}-\mu_d
=(\beta-\rho_c+o(1))d\ge\gamma d
\end{align}
for all sufficiently large $N$. Hence,
\begin{align}
\Pr(U>\beta q/c)
&\le\Pr(U-\mu_d>\gamma d)
\nonumber\\
&=o\left(\frac{1}{N}\right).
\label{eq:sharp-root-tail}
\end{align}
Observe that the root union under consideration is \(R(h)\), where \(h=\prod_{j=1}^d F_j\). Applying Lemma~\ref{lem:conditional-covering}(ii) with $L=\beta q/c<q$, we obtain
\begin{align}
\Pr(\Psi\subseteq R(h),\ U\le\beta q/c)
\le\left(\frac{\beta}{c}\right)^n.
\label{eq:sharp-cover-small}
\end{align}
Splitting the event in the definition \eqref{eq:dd-pm} according to whether $U\le\beta q/c$, and using \eqref{eq:sharp-root-tail} and \eqref{eq:sharp-cover-small}, yields
\begin{align}
p_d\le\left(\frac{\beta}{c}\right)^n+o\left(\frac1N\right).
\end{align}
The tail term is independent of $\Psi$ and $n$, so the bound holds throughout the range $1\le n\le\lfloor\beta d\rfloor$ stated in Lemma~\ref{lem:root-covering}.

\section{Proof of Theorem~\ref{thm:sharpbound} (Main Result for COMP Decoding)}\label{app: proofSharpBound}

Fix $\varepsilon>0$. Recall from Appendix~\ref{app: proofthmwarmup} the initial upper bound \eqref{eq:warmup_union} on $P_e$ and the polynomial-translation equivalence \eqref{eq:cover-root-equivalence}. For a fixed nondefective item $i$, subtracting $f_i$ from the defective polynomials gives a uniform $d$-subset of $\mathcal P_{<k}^*$, whose roots cover $\Psi$ exactly when item $i$ is covered. Its covering probability is therefore $p_d$ defined in \eqref{eq:dd-pm}, and \eqref{eq:warmup_union} becomes
\begin{align}
P_e\le(N-d)p_d.
\label{eq:comp-pd}
\end{align}

We choose a fixed limit $c>0$ and $\gamma\in(0,c-\rho_c)$, take an admissible sequence with $q/d\to c$ (Definition \ref{def:admissible}), and set
\begin{align}
\beta:=\rho_c+\gamma.
\label{eq:sharp-beta}
\end{align}
Let $\delta>0$ be fixed and choose
\begin{align}
	n = \left\lceil
	\frac{(1+\delta)\log N}
	{\log(c/\beta)}
	\right\rceil.
	\label{eq:sharp-n}
\end{align}
We claim that this choice satisfies the conditions of Lemma~\ref{lem:root-covering}, whose statement then gives
\begin{align}
	p_d\le\left(\frac{\beta}{c}\right)^n
	+o\left(\frac{1}{N}\right).
	\label{eq:sharp-Ai}
\end{align}
We proceed with the required verification as follows.  
Since $\beta<c$, the denominator in \eqref{eq:sharp-n} is a positive constant, so $n=\Theta(\log N)$. Together with $d=\Omega(\log^2N)$ and $q=\Theta(d)$, this gives $n=o(d)$, $n\le\lfloor\beta d\rfloor$, and $n\le q$ eventually. Moreover, $k=\log N/\log q$ gives $n/k=\Theta(\log q)\to\infty$, ensuring $k\le n$. Hence, the range in Lemma~\ref{lem:root-covering} and the Reed--Solomon constraints both hold. 

Since $\beta/c<1$, the choice of $n$ in \eqref{eq:sharp-n} gives
\begin{align}
\left(
\frac{\beta}{c}
\right)^n
\le
N^{-(1+\delta)},
\end{align}
and substitution into \eqref{eq:sharp-Ai} yields
\begin{align}
p_d
\le
N^{-(1+\delta)}
+
o\left(
\frac{1}{N}
\right).
\end{align}
Combining with \eqref{eq:comp-pd} then gives
\begin{align}
P_e
&\le
N^{-\delta}
+
o(1)
\longrightarrow0.
\label{eq:sharp-Pe}
\end{align}

It remains to choose the parameters to optimize the number of tests. Since $T=nq$ and $q=(c+o(1))d$, \eqref{eq:sharp-n} yields
\begin{align}
T
=
\left(
\frac{c(1+\delta)}
{\log(c/\beta)}
+
o(1)
\right)
d\log N.
\label{eq:sharp-tests-gamma}
\end{align}
The constants $c,\beta,\delta$ are fixed before taking $N\to\infty$. We proceed to optimize the resulting coefficient: For fixed $c$, letting $\gamma\downarrow0$ gives $\beta\downarrow\rho_c$. Subsequently letting $\delta\downarrow0$, the limiting coefficient becomes
\begin{align}
C(c)
&:=
\frac{c}
{\log(c/\rho_c)}
\nonumber\\
&=
\frac{c}
{\log\left(
\frac{1}{1-e^{-1/c}}
\right)}.
\label{eq:sharp-C}
\end{align}

We now minimize $C(c)$ over $c>0$. Set
\begin{align}
x:=\frac{1}{c}.
\end{align} 
Since $\log$ is base $2$,
\begin{align}
C(c)
=
\frac{\ln2}
{x\left[-\ln(1-e^{-x})\right]}.
\label{eq:sharp-C-x}
\end{align}
Hence, minimizing $C(c)$ is equivalent to maximizing
\begin{align}
g(x)
:=
x\left[-\ln(1-e^{-x})\right],
\qquad
x>0.
\end{align}
Differentiating gives
\begin{align}
g'(x)
=
-\ln(1-e^{-x})
-
\frac{x}{e^x-1}.
\end{align}
At $x=\ln2$, we have
\begin{align}
-\ln(1-e^{-x})
=\ln2,
\qquad
\frac{x}{e^x-1}
= \ln2,
\end{align}
so $g'(\ln2)=0$.  Moreover, the expression for $g'(x)$ shows that it is positive for $x<\ln 2$ and negative for $x>\ln 2$, so that $x=\ln 2$ is the unique global maximizer


Hence, the optimum is
\begin{align}
c_\star
=
\frac{1}{\ln2},
\end{align}
and under this choice, we have
\begin{align}
e^{-1/c_\star}
=
\frac{1}{2},
\qquad
\rho_{c_\star}
=
\frac{c_\star}{2},
\end{align}
and consequently
\begin{align}
\log\left(
\frac{c_\star}{\rho_{c_\star}}
\right)
=
\log2
=
1.
\end{align}
It follows that
\begin{align}
\inf_{c>0}C(c)
=
\frac{1}{\ln2}.
\label{eq:sharp-optimum}
\end{align}

Finally, we choose admissible parameters with limit $c=c_\star=1/\ln2$. Because the coefficient in \eqref{eq:sharp-tests-gamma} converges to $1/\ln2$ as $\gamma\downarrow0$ and $\delta\downarrow0$, we may choose fixed $\gamma,\delta>0$ sufficiently small so that
\begin{align}
\frac{c(1+\delta)}
{\log(c/\beta)}
<
\frac{1}{\ln2}
+
\frac{\varepsilon}{2}.
\end{align}
For all sufficiently large $N$, the $o(1)$ term in \eqref{eq:sharp-tests-gamma} is smaller than $\varepsilon/2$. Therefore,
\begin{align}
T
\le
\left(
\frac{1}{\ln2}
+
\varepsilon
\right)
d\log N.
\end{align}
Together with \eqref{eq:sharp-Pe}, this proves the theorem.

\section{Proof of Lemma~\ref{lem:alpha-threshold} (Asymptotic Limit of the Distinct Root Count)}\label{app: proofAlpha}

For the first statement, the mean part of Lemma~\ref{lem:betterbound} immediately gives $\mu_d/d\to\rho_c$ (without requiring $d=\Omega(\log^2N)$). It remains to prove concentration under $k^2=o(d)$. 
Let
\begin{align}
Z_i
:=
\mathbb E
\left[
U
\,\middle|\,
F_1,\ldots,F_i
\right],
\qquad
i=0,1,\ldots,d.
\end{align}
Then $(Z_i)_{i=0}^d$ is the Doob martingale associated with revealing the sampled polynomials sequentially, with
\begin{align}
Z_0
=
\mathbb EU,
\qquad
Z_d
=
U.
\end{align}

We claim that, conditioned on $F_1,\ldots,F_{i-1}$,
the range of $Z_i$ has length at most $k-1$.
To see this, fix the previously revealed polynomials and two
distinct remaining values $f,g$ of $F_i$.
Given $F_i=f$, sample the subsequent polynomials uniformly
without replacement from the unused population excluding $f$.
We couple this with the case $F_i=g$ by replacing $g$ by $f$
among the subsequent polynomials if $g$ appears, and leaving
them unchanged otherwise.
This gives a bijection between the two sets of possible
outcomes, preserving their uniform distributions.  
Under this coupling, the two collections of sampled
polynomials are either identical or differ only by replacing
$f$ with $g$.  
In the latter case, let $V$ denote the union of the root sets
of the common polynomials, and let $U_f$ and $U_g$ denote
the respective numbers of distinct roots. Then, we have
\begin{align}
|U_f-U_g|
&= \bigl||R(f)\setminus V|-|R(g)\setminus V|\bigr| \notag\\
&\le \max\{|R(f)|,|R(g)|\} \notag\\
&\le k-1.
\end{align}
Since $|U_f-U_g|\le k-1$ pointwise, their conditional expectations differ by at most $k-1$. This bounds the conditional range of $Z_i$ by $k-1$, as claimed.

The Azuma--Hoeffding inequality for martingales with conditional range lengths bounded by $k-1$ consequently gives, for every $t\ge0$,
\begin{align}
\Pr
\left(
\left|
U-\mathbb EU
\right|
\ge t
\right)
\le
2\exp\left(
-\frac{2t^2}
{d(k-1)^2}
\right).
\label{eq:U-Azuma}
\end{align}
Fix any $\xi>0$ and set $t=\xi d$. Then
\begin{align}
\Pr
\left(
\left|
\frac{U}{d}
-
\frac{\mathbb EU}{d}
\right|
\ge\xi
\right)
\le
2\exp\left(
-\frac{2\xi^2d}
{(k-1)^2}
\right).
\end{align}
By the assumption $k^2=o(d)$, we have
\begin{align}
\frac{d}{(k-1)^2}
\longrightarrow\infty,
\end{align}
and hence
\begin{align}
\frac{U-\mathbb EU}{d}
\xrightarrow{\mathrm{P}}
0.
\label{eq:U-centered-convergence}
\end{align}
Combining \eqref{eq:union-mean-limit} and \eqref{eq:U-centered-convergence} yields
\begin{align}
\frac{U}{d}
\xrightarrow{\mathrm{P}}
\rho_c.
\end{align}

Finally, fix any $\eta<\rho_c$ and let
\begin{align}
\xi
:=
\frac{\rho_c-\eta}{2}
>0.
\end{align}
Then
\begin{align}
\rho_c-\xi
=
\frac{\rho_c+\eta}{2}
>
\eta.
\end{align}
Therefore,
\begin{align}
\Pr
\left(
U>\eta d
\right)
&=
\Pr
\left(
\frac{U}{d}>\eta
\right)
\nonumber\\
&\ge
\Pr
\left(
\frac{U}{d}>\rho_c-\xi
\right)
\nonumber\\
&\ge
1-
\Pr
\left(
\left|
\frac{U}{d}-\rho_c
\right|
\ge\xi
\right).
\end{align}
Since $U/d\xrightarrow{\mathrm{P}}\rho_c$, the final probability converges to zero, giving
\begin{align}
\Pr
\left(
U>\eta d
\right)
\longrightarrow1.
\end{align}
This completes the proof.

\section{Proof of Theorem~\ref{thm:dd-sharpbound} (Main Result for DD Decoding)}\label{app: proofDD}

Recall that each item $j$ corresponds to a polynomial
$f_j\in\mathcal P_{<k}$, and participates in test $(a,y)$
if and only if $f_j(a)=y$, where $a\in\Psi$.
We also recall the root-covering probabilities $p_m$
defined in \eqref{eq:dd-pm}.

Let $S\subseteq[N]$ be a uniformly random defective set of cardinality $d$. For a fixed defective set $s$, define
\begin{align}
C(s)
:=
\left\{
j\in[N]\setminus s:
\operatorname{supp}(A_j)
\subseteq
\bigcup_{i\in s}
\operatorname{supp}(A_i)
\right\}.
\label{eq:dd-covered-set}
\end{align}
After eliminating every item appearing in a negative test, the set of possible defectives is exactly
\begin{align}
s\cup C(s).
\end{align}

For $i\in s$, define 
\begin{align}
P_{f_i}^{\Psi}(s\setminus\{i\})
:=
\left\{
a\in\Psi:
f_j(a)\neq f_i(a)\ \forall j\in s\setminus\{i\}
\right\},
\label{eq:dd-private}
\end{align}
which represents the set of ``private'' evaluation points of $f_i$ relative to the remaining defectives (i.e., coordinates where $f_i$ differs from every other defective polynomial).

Suppose that $C(s)=\varnothing$ and that
\begin{align}
P_{f_i}^{\Psi}(s\setminus\{i\})
\neq
\varnothing
\end{align}
for every $i\in s$. Fix $i\in s$ and choose
\begin{align}
a_i
\in
P_{f_i}^{\Psi}(s\setminus\{i\}).
\end{align}
Then the positive test $(a_i,f_i(a_i))$ contains item $i$ and no other defective item. Since $C(s)=\varnothing$, the possible-defective set is precisely $s$, and therefore $i$ is the unique possible defective contained in this test. DD consequently declares $i$ defective. Since this holds for every $i\in s$, DD recovers $s$ exactly.

It follows that the DD error probability satisfies
\begin{align}
P_e
&\le
\Pr\left(
C(S)\neq\varnothing
\right)
\nonumber\\
&\quad+
\Pr\left(
\exists i\in S:
P_{f_i}^{\Psi}(S\setminus\{i\})
=
\varnothing
\right).
\label{eq:dd-failure-decomposition}
\end{align}
We will show that both terms in \eqref{eq:dd-failure-decomposition} are controlled by the same root-covering probability.

We first show a monotonicity property.

\begin{lemma}
\label{lem:dd-pm-monotone}
The sequence $(p_m)_{1\le m\le M}$ is nondecreasing. In particular,
\begin{align}
p_{d-1}\le p_d.
\end{align}
\end{lemma}

\begin{proof}
For $2\le m\le M$, sample
\begin{align}
B_m
\sim
\operatorname{Unif}
\left(
\binom{\mathcal P_{<k}^*}{m}
\right)
\end{align}
and remove one element uniformly from $B_m$. The resulting set $B_{m-1}$ is uniformly distributed over all $(m-1)$-subsets of $\mathcal P_{<k}^*$. Moreover,
\begin{align}
\bigcup_{h\in B_{m-1}}R(h)
\subseteq
\bigcup_{h\in B_m}R(h).
\end{align}
Hence the root-covering event for $B_{m-1}$ implies the corresponding event for $B_m$, which gives $p_{m-1}\le p_m$.
\end{proof}

We next bound the first term of \eqref{eq:dd-failure-decomposition}. For a fixed $j\in[N]$, condition on $j\notin S$ and recall the nondefective-item translation argument from Appendix~\ref{app: proofthmwarmup} (and recapped in Appendix~\ref{app: proofSharpBound}). Subtracting $f_j$ gives a uniform $d$-subset of $\mathcal P_{<k}^*$, and its roots cover $\Psi$ exactly when $j\in C(S)$. Hence, by the definition \eqref{eq:dd-pm},
\begin{align}
\Pr\left(j\in C(S)\,\middle|\,j\notin S\right)=p_d.
\end{align}
Linearity of expectation then gives
\begin{align}
\mathbb E|C(S)|
&=
\sum_{j=1}^N
\Pr\left(
j\in C(S)
\right)
\nonumber\\
&=
\sum_{j=1}^N
\Pr(j\notin S)
\Pr\left(
j\in C(S)
\,\middle|\,
j\notin S
\right)
\nonumber\\
&=
N\frac{N-d}{N}p_d
\nonumber\\
&=
(N-d)p_d.
\end{align}
Since $|C(S)|$ is nonnegative and integer-valued, it follows that
\begin{align}
\Pr\left(
C(S)\neq\varnothing
\right)
&=
\Pr\left(
|C(S)|\ge1
\right)
\nonumber\\
&\le
\mathbb E|C(S)|
\nonumber\\
&=
(N-d)p_d.
\label{eq:dd-covered-probability}
\end{align}

We now consider the second term of \eqref{eq:dd-failure-decomposition}. Fix $i\in[N]$ and condition on $i\in S$. Then
\begin{align}
S\setminus\{i\}
\sim
\operatorname{Unif}
\left(
\binom{[N]\setminus\{i\}}{d-1}
\right).
\end{align}
Again, translation by $-f_i$ is a bijection from $\mathcal P_{<k}\setminus\{f_i\}$ onto $\mathcal P_{<k}^*$. Hence,
\begin{align}
B_i
:=
\left\{
f_j-f_i:
j\in S\setminus\{i\}
\right\}
\end{align}
is a uniformly random $(d-1)$-subset of $\mathcal P_{<k}^*$.

By the definition in \eqref{eq:dd-private}, the condition 
$P_{f_i}^{\Psi}(S\setminus\{i\})=\varnothing$ holds
if and only if for every $a\in\Psi$ there exists some $j\in S\setminus\{i\}$ such that $f_j(a)=f_i(a)$. Equivalently,
\begin{align}
P_{f_i}^{\Psi}(S\setminus\{i\})
=
\varnothing
\quad\Longleftrightarrow\quad
\Psi
\subseteq
\bigcup_{h\in B_i}R(h).
\end{align}
Consequently,
\begin{align}
&\Pr\left(
P_{f_i}^{\Psi}(S\setminus\{i\})
=
\varnothing
\,\middle|\,
i\in S
\right)
\nonumber\\
&=
p_{d-1}
\le
p_d,
\label{eq:dd-private-pd}
\end{align}
where the final inequality follows from Lemma~\ref{lem:dd-pm-monotone}.

A union bound over the defective items now gives
\begin{align}
&\Pr\left(
\exists i\in S:
P_{f_i}^{\Psi}(S\setminus\{i\})
=
\varnothing
\right)
\nonumber\\
&\le
\sum_{i=1}^N
\Pr\left(
i\in S,\,
P_{f_i}^{\Psi}(S\setminus\{i\})
=
\varnothing
\right)
\nonumber\\
&=
\sum_{i=1}^N
\Pr(i\in S)
\Pr\left(
P_{f_i}^{\Psi}(S\setminus\{i\})
=
\varnothing
\,\middle|\,
i\in S
\right)
\nonumber\\
&\le
N\frac{d}{N}p_d
\nonumber\\
&=
dp_d.
\label{eq:dd-private-probability}
\end{align}

Combining \eqref{eq:dd-failure-decomposition}, \eqref{eq:dd-covered-probability}, and \eqref{eq:dd-private-probability}, we obtain
\begin{align}
P_e
\le
(N-d)p_d+dp_d
=
Np_d.
\label{eq:dd-Pe-pd}
\end{align}

Fix the same constants $c$, $\beta\in(\rho_c,c)$, and $\delta>0$ as in Appendix~\ref{app: proofSharpBound}, and choose $n$ by \eqref{eq:sharp-n}. Lemma~\ref{lem:root-covering} and \eqref{eq:dd-Pe-pd} give
\begin{align}
P_e
&\le N\left(\frac{\beta}{c}\right)^n
+o(1)
\nonumber\\
&\le N^{-\delta}+o(1)
\longrightarrow0.
\label{eq:dd-error-final}
\end{align}
The test count is exactly the one in \eqref{eq:sharp-tests-gamma}. The fixed-parameter choices and optimization in \eqref{eq:sharp-C}--\eqref{eq:sharp-optimum} therefore give $T\le(1/\ln2+\varepsilon)d\log N$ for every fixed $\varepsilon>0$. This proves DD at the COMP-level threshold; it does not attain the smaller random-design DD constant for a fixed positive sparsity exponent.

\end{document}